\documentclass[aps,12pt,tightenlines,amsmath,amssymb,showpacs]{revtex4}
\usepackage{graphicx}
\usepackage{dcolumn}
\usepackage{bm}

\def\Eqref#1{Eq.~(\ref{#1})}
\newcommand{\be}{\begin{equation}}\newcommand{\ee}{\end{equation}}
\newcommand{\bea}{\begin{eqnarray}}\newcommand{\eea}{\end{eqnarray}}
\newcommand{\bean}{\begin{eqnarray*}}\newcommand{\eean}{\end{eqnarray*}}

\begin{document}

\title{QUANTUM HAMILTON-JACOBI THEORY}

\author{Marco Roncadelli}
\affiliation{INFN, Sezione di Pavia, Via A. Bassi 6, I-27100 Pavia, Italy, and Dipartimento di Fisica Nucleare e Teorica, Universit\`a di Pavia, Italy}
\email{marco.roncadelli@pv.infn.it}

\author{L. S. Schulman}
\affiliation{Physics Department, Clarkson University, Potsdam, New York 13699-5820, USA}
\email{schulman@clarkson.edu}

\date{\today}
\begin{abstract}
Quantum canonical transformations have attracted interest since the beginning of quantum theory. Based on their classical analogues, one would expect them to provide a powerful quantum tool. However, the difficulty of solving a nonlinear operator partial differential equation such as the quantum Hamilton-Jacobi equation (QHJE) has hindered progress along this otherwise promising avenue. We overcome this difficulty. We show that solutions to the QHJE can be constructed by a simple prescription starting from the propagator of the associated Schr\"odinger equation. Our result opens the possibility of practical use of quantum Hamilton-Jacobi theory. As an application we develop a surprising relation between operator ordering and the density of paths around a semiclassical trajectory.
\end{abstract}

\maketitle


Canonical transformations play a central role in classical mechanics~\cite{Goldstein}. From the earliest days of quantum mechanics, the importance of \textit{quantum} canonical transformations (QCT's) has been recognized~\cite{Born} and their properties have been systematically investigated by Jordan~\cite{Jordan}, London~\cite{London}, Dirac~\cite{Dirac} and Schwinger~\cite{Schwinger1}, among others. Schwinger's framework---based on the Quantum Action Principle~\cite{Schwinger2}---provides the most suitable context to define a QCT as $\hat q \to \hat Q, \hat p \to \hat P, H(\hat q, \hat p,t) \to K(\hat Q, \hat P,t)$, where all canonical variables pertain to the same dynamical system $\cal S$ with $N$ degrees of freedom~\cite{hat}.

Owing to the formal similarities between classical and quantum mechanics, QCT's closely resemble their classical counterparts. In particular, one of four possible sets of independent canonical variables $(\hat q, \hat Q)$, $(\hat q, \hat P)$, $(\hat Q, \hat p)$, $(\hat p, \hat P)$ must be selected to represent a QCT explicitly, and we denote by $W(\hat q, \hat Q, t)$, $W(\hat q, \hat P, t)$, etc., the associated operator generating functions. Choosing the set $(\hat q, \hat Q)$, a QCT can be written as ($1 \leq i \leq N$)
\begin{equation}
\hat p_i = \frac{\partial}{\partial \hat q_i} W(\hat q, \hat Q, t)~,
\label{a1}
\end{equation}
\begin{equation}
\hat P_i = - \frac{\partial}{\partial \hat Q_i} W(\hat q, \hat Q, t)~, 
\label{a2}
\end{equation}
\begin{equation}
K(\hat Q, \hat P,t) = H(\hat q, \hat p,t) + \frac{\partial}{\partial t} W(\hat q, \hat Q, t)\;.
\label{a3}
\end{equation}

The presence of noncommuting operators in the generating function makes QCT's different from the classical ones and is ultimately responsible for the difference between classical and quantum mechanics~\cite{Sudarshan}. As emphasized by Jordan and Dirac, the resulting operator-order ambiguity should be fixed by enforcing {\it well-ordering\/}: operators represented by capital letters should always stay to the right of those labelled by lower case letters. This means that $W(\hat q, \hat Q, t)$ should have the structure
\begin{equation}
W(\hat q, \hat Q, t) = \sum_{\alpha} f_{\alpha} (\hat q, t) \, g_{\alpha} (\hat Q, t)~,
\label{a4}
\end{equation}
for suitable functions $f_{\alpha}(\cdot)$ and $g_{\alpha}(\cdot)$. Throughout, we will suppose that operator generating functions are {\it well-ordered}. Note that with {\it well-ordering\/} a quantum generating function like $W(\hat q, \hat Q, t)$ is uniquely defined by the replacements $q \to \hat q$, $Q \to \hat Q$ in a given c-number function $W(q, Q, t)$~\cite{cnumber}.

As in classical mechanics, the quantum time evolution is described by a canonical transformation bringing the canonical variables in the Heisenberg picture $\hat q(t)$, $\hat p(t)$ to constant values at some initial time $t_0$. In addition, $\hat q(t)$, $\hat p(t)$ can be derived from Eqs.\ (\ref{a1}) and (\ref{a2}), provided that the transformed Hamiltonian vanishes. As a consequence, the operator generating function $W(\hat q, \hat Q, t)$ obeys the operator {\it quantum Hamilton-Jacobi equation} (QHJE)
\begin{equation}
H \left( \hat q, \frac{\partial}{\partial \hat q} W(\hat q, \hat Q, t), t \right) +
\frac{\partial}{\partial t} W(\hat q, \hat Q, t) = 0~.
\label{a5}
\end{equation}
$W(\hat q, \hat Q, t)$ should be a complete solution of \Eqref{a5}, i.e., it should depend on $N$ independent ``integration constants'' ${\hat Q}_i$. As in classical mechanics, the operator Hamilton-Jacobi equation, \Eqref{a5}, provides an independent formulation of the theory. Yet the formidable difficulty of finding solutions to this nonlinear operator partial differential equation has hindered progress along this otherwise promising avenue.

Our aim is to show that this stumbling block can be sidestepped, thereby opening the way to exploiting the operator QHJE as a calculational tool. As we will demonstrate, the solutions to the operator QHJE arise by a simple prescription from the solutions of the Schr\"odinger equation for the same Hamiltonian. In particular, the operator generating function $W(\hat q, \hat Q, t)$ arises from the quantum propagator. Implications of our result will be discussed after we have completed its demonstration.

We are concerned throughout with the general {\it Weyl-ordered} 
Hamiltonian~\cite{TDLee}
\begin{equation}
\vbox{\displaylines{H(\hat q, \hat p, t) = \frac{1}{2} a_{ij}(\hat q) \hat p_i \hat p_j + \hat p_i a_{ij}(\hat q) \hat p_j + \frac{1}{2} \hat p_i \hat p_j a_{ij}(\hat q) \cr + b_i(\hat q) \hat p_i + \hat p_i b_i(\hat q) +
c(\hat q)~,\cr}}
\label{a6}
\end{equation}
where $a_{ij}(\cdot)$, $b_i(\cdot)$, and $c(\cdot)$ are functions of ${\hat q}_k$, and summation over repeated Latin indices for the degrees of freedom of $\cal S$ is understood. Employing the shorthand $\hat W \equiv W(\hat q, \hat Q, t)$, \Eqref{a5} reads
\bea
\frac12 a_{ij}(\hat q) \frac{\partial\hat W}{\partial\hat q_i}
                      \frac{\partial\hat W}{\partial \hat q_j} 
&+& 
\frac{\partial \hat W}{\partial \hat q_i} a_{ij}(\hat q)
               \frac{\partial \hat W}{\partial \hat q_j} 
+ 
\frac{1}{2} \frac{\partial \hat W}{\partial \hat q_i}
\frac{\partial \hat W}{\partial \hat q_j} a_{ij}(\hat q) 
+ 
b_i(\hat q) \frac{\partial \hat W}{\partial \hat q_i}
\nonumber\\
&+& 
\frac{\partial \hat W}{\partial \hat q_i} b_i(\hat q) 
+ c(\hat q) + \frac{\partial \hat W}{\partial t} 
=0~.
\label{a7}
\eea
Since we are looking for the relationship between the operator QHJE and the Schr\"odinger equation, we turn \Eqref{a7} into a c-number partial differential equation. Hence we sandwich \Eqref{a7} between $\langle q|$ and $|Q \rangle$, finding
\bea
\frac{1}{2} a_{ij}(q) \langle q \left| \frac{\partial \hat W}{\partial \hat q_i}
\frac{\partial \hat W}{\partial \hat q_j} \right|Q \rangle 
+ 
\langle q \left| \frac{\partial \hat W}{\partial \hat q_i} a_{ij}(\hat q) \frac{\partial \hat W}{\partial \hat q_j} \right| Q \rangle 
&&\nonumber\\ 
+
\frac{1}{2} \langle q \left|
\frac{\partial \hat W}{\partial \hat q_i} \frac{\partial \hat W}{\partial \hat q_j} a_{ij}(\hat q) \right| Q \rangle
+ 
b_i(q) \langle q \left| \frac{\partial \hat W}{\partial \hat q_i} \right|Q \rangle 
&&\nonumber\\ 
+
\langle q \left| \frac{\partial \hat W}{\partial \hat q_i} b_i(\hat q) \right| Q \rangle 
+
c(q) \langle q| Q \rangle 
+
\langle q \left| \frac{\partial \hat W}{\partial t} \right|Q \rangle &&= 0   \,.
\label{a8}
\eea
To evaluate the matrix elements in \Eqref{a8}, we make repeated use of the canonical commutation relations. In this connection, we recall that for an arbitrary function $G(\cdot)$
\begin{equation}
[G(\hat q), \hat p_i] = i \hbar \frac{\partial G(\hat q)}{\partial \hat q_i}~.
\label{a9}
\end{equation}
By inserting \Eqref{a1} into (\ref{a9}), we obtain
\begin{equation}
\frac{\partial \hat W}{\partial \hat q_i} G(\hat q) = G(\hat q) \frac{\partial \hat W}{\partial \hat q_i} -
i \hbar \frac{\partial G(\hat q)}{\partial \hat q_i}~.
\label{a10}
\end{equation}
We begin by taking $G(\hat q) \equiv b_i(\hat q)$, $G(\hat q) \equiv a_{ij}(\hat q)$ and $G(\hat q) \equiv \partial a_{ij}(\hat q)/ \partial {\hat q}_j$. Accordingly, Eq. (\ref{a10}) allows us to rewrite \Eqref{a8} as
\bea
2 a_{ij}(q) \langle q \left| \frac{\partial \hat W}{\partial \hat q_i}
\frac{\partial \hat W}{\partial \hat q_j} \right|Q \rangle + 2 \left( b_i(q) - i \hbar \frac{\partial
a_{ij}(q)}{\partial q_j} \right) \langle q \left| \frac{\partial \hat W}{\partial \hat q_i} \right|Q \rangle &&
\nonumber\\ 
+
\left( c(q) -i \hbar \frac{\partial b_i(q)}{\partial q_i} - \frac{{\hbar}^2}{2}
\frac{\partial^2 a_{ij}(q)}{\partial q_i \partial q_j} \right) \langle q |Q \rangle +
\langle q \left| \frac{\partial \hat W}{\partial t} \right|Q \rangle
&&
\!\!\!\!
= 0~.
\label{a11}
\eea
At this point, we denote by $W(q, Q, t)$ the c-number function that uniquely produces $W(\hat q, \hat Q, t)$ by the substitution $q \to \hat q$, $Q \to \hat Q$. Explicit use of
\Eqref{a4} yields
\begin{equation}
\langle q| {\hat W} |Q \rangle = W(q,Q,t) \langle q|Q \rangle \,,
\label{a12k}
\end{equation}
\begin{equation}
\left. \right. \langle q \left| \frac{\partial \hat W}{\partial t} \right|Q \rangle = \frac{\partial W(q,Q,t)}{\partial t} \langle q|Q \rangle \,,
\label{a12}
\end{equation}
\begin{equation}
\left. \right. \langle q \left| \frac{\partial \hat W}{\partial \hat q_i} \right|Q \rangle =
\frac{\partial W(q,Q,t)}{\partial q_i} \langle q|Q \rangle \,,
\label{a13}
\end{equation}
and furthermore
\begin{equation}
\left. \right. \langle q \left| \frac{\partial \hat W}{\partial \hat q_i}
\frac{\partial \hat W}{\partial \hat q_j} \right|Q \rangle = \sum_{{\alpha},{\beta}} \langle q \left|
\frac{\partial f_{\alpha} (\hat q, t)}{\partial \hat q_i} g_{\alpha} (\hat Q, t) \frac{\partial f_{\beta} (\hat q, t)}{\partial \hat q_j} g_{\beta} (\hat Q, t) \right|Q \rangle~.
\label{a14}
\end{equation}
What remains to be done is to disentangle \Eqref{a14}. To this end, we first take $G(\hat q) \equiv \partial f_{\beta}(\hat q, t)/ \partial \hat q_j$ in \Eqref{a10} to get
\bea
\sum_{\alpha} \frac{\partial f_{\alpha} (\hat q, t)}{\partial \hat q_i}
g_{\alpha} (\hat Q, t) \frac{\partial f_{\beta}
(\hat q, t)}{\partial \hat q_j} 
&=&
\frac{\partial f_{\beta}(\hat q, t)}{\partial \hat q_j}
      \sum_{\alpha} \frac{\partial f_{\alpha} (\hat q, t)}{\partial \hat q_i} g_{\alpha} (\hat Q, t) 
\nonumber \\
&&
- i \hbar \frac{\partial^2 f_{\beta} (\hat q,t)}{\partial \hat q_i \partial \hat q_j}~.
\label{a15}
\eea
We next multiply \Eqref{a15} by $g_{\beta}(\hat Q, t)$ on the right and sum over $\alpha$, thereby obtaining
\bea
\sum_{{\alpha},{\beta}} \frac{\partial f_{\alpha} (\hat q, t)}{\partial \hat q_i} g_{\alpha} (\hat Q, t) \frac{\partial f_{\beta}
(\hat q, t)}{\partial \hat q_j} g_{\beta} (\hat Q, t) 
&&
\nonumber \\ 
= 
\sum_{{\alpha},{\beta}} \frac{\partial f_{\beta} (\hat q, t)}{\partial \hat q_j} \frac{\partial f_{\alpha} (\hat q, t)}{\partial \hat q_i} g_{\alpha} (\hat Q, t) g_{\beta} (\hat Q, t)  
&&
\!\!\!\!
- i \hbar \sum_{\beta}
\frac{\partial^2 f_{\beta} (\hat q,t)}{\partial \hat q_i \partial \hat q_j} g_{\beta} (\hat Q, t)~,
\label{a16}
\eea
which allows us to rewrite \Eqref{a14} as
\begin{equation}
\langle q \left| \frac{\partial \hat W}{\partial \hat q_i}
\frac{\partial \hat W}{\partial \hat q_j} \right|Q \rangle 
= \left( \frac{\partial W(q,Q,t)}{\partial q_i} \frac{\partial W(q,Q,t)}{\partial q_j} - i \hbar \frac{\partial^2 W(q,Q,t)}{\partial q_i \partial q_j} \right) \langle q | Q \rangle \;.
\label{a17}
\end{equation}
As a consequence, \Eqref{a11} takes the form
\bea
2 a_{ij}(q)
\left(\frac{\partial W(q,Q,t)}{\partial q_i} \frac{\partial W(q,Q,t)}{\partial q_j}
      - i \hbar \frac{\partial^2 W(q,Q,t)}{\partial q_i \partial q_j} \right) 
&&
\nonumber\\
+ 2 \left( b_i(q) - i \hbar \frac{\partial a_{ij}(q)}{\partial q_j} \right)
     \frac{\partial W(q,Q,t)}{\partial q_i}
+ c(q) -i \hbar \frac{\partial b_i(q)}{\partial q_i} 
&&
\nonumber\\
- \frac{{\hbar}^2}{2}
\frac{\partial^2 a_{ij}(q)}{\partial q_i \partial q_j}
 + \frac{\partial W(q,Q,t)}{\partial t}
&& 
\!\!\!\!
= 0~.
\label{a18}
\eea
This derivation makes it natural to regard \Eqref{a18} as the c-number QHJE associated with the
operator QHJE (\ref{a5}) for $\cal S$ described by the quantum Hamiltonian (\ref{a6}).

The physical significance of \Eqref{a18} becomes clear by setting
\begin{equation}
\psi (q,Q,t) \equiv {\rm exp} \{(i/{\hbar}) W(q, Q,t) \} \;.
\label{a19}
\end{equation}
A straightforward (if tedious) calculation shows that $\psi (q,Q,t)$ obeys precisely the 
Schr\"odinger equation associated with the quantum Hamiltonian (\ref{a6}) in the variables 
$q,t$~\cite{Gottfried}. Hence -- thanks to Eqs. (\ref{a12k}) and (\ref{a19}) -- starting from a solution $W(\hat q, \hat Q, t)$ we get a solution $\psi (q,Q,t)$ of the corresponding 
Schr\"odinger equation depending on $N$ independent constants $Q_i$. We stress that this 
result holds true even for solutions $W( \hat q,t)$ of the operator QHJE that are {\it independent} of $\hat Q$, since all equations from (\ref{a7}) onward could have been multiplied by $\int dQ \, \phi(Q)$, with $\phi (Q)$ arbitrary. What is more important for us, the argument can be turned around, because $W(\hat q, \hat Q, t)$ can be {\it uniquely} obtained from $W(q, Q, t)$ by enforcing {\it well-ordering}. Therefore, from a solution $\psi (q,Q,t)$ of the Schr\"odinger equation  depending on $N$ independent constants $Q_i$ we get $W(\hat q, \hat Q, t)$, and from any particular solution 
$\psi (q,t)$ we can construct a particular solution $W(\hat q, t)$.

So far, we have focused on showing that $W(q,Q,t)$ satisfies a certain differential equation. As we demonstrate below, by use of appropriate boundary conditions we get more specific information. Namely, the solution of Schr\"odinger's equation that results from the operator generating function $W(\hat q, \hat Q, t)$ is precisely the quantum propagator $K(q,Q,t)$~\cite{Schulman}. Since any solution of the Schr\"odinger equation arises by convolving an arbitrary wave function with the propagator, we conclude that {\it any} solution of the operator QHJE can ultimately be constructed in terms of the propagator.

This will allow solutions of and approximations to the operator QHJE to be obtained, since a wealth of information is available on the corresponding solutions to Schr\"odinger's equation. In particular, once an exact or approximate $\hat W$ has been constructed, one can obtain the time dependence of operators, using Eqs.\ (\ref{a1}) and~(\ref{a2}).

We proceed to prove that $\psi (q,Q,t) = K(q,Q,t)$. Since both quantities satisfy the same 
Schr\"odinger equation, which is first order in time, all we need show is that they have the same boundary conditions at $t=0$. The propagator of course is $\delta(q-Q)$ at $t = 0$. To show that $\psi(q,Q,t)$ shares this property, we must look at the behavior of $\hat W$ for $t\to0$. We expect $\hat W$ to generate the identity transformation in the limit $t\to0$, but there is a slight complication: As in classical mechanics \cite{Goldstein}, the identity transformation using the $(\hat q,\hat Q)$ variables does not have a simple form.

A way out of this difficulty relies on the observation that for a nonsingular potential the solution to the classical Hamilton-Jacobi equation approaches that of the free particle for $t\to0$. Thus, for sufficiently small $t$ the classical generating function has the form $F(q,Q)=m(Q-q)^2/2t$. We use this to guess the limit of the \textit{operator} $\hat W$ for $t\to0$, and from that to obtain the corresponding limit of the c-number function $W(q,Q,t)$. The first observation is that as a candidate for the small-$t$ limit of $\hat W$, the {\it well-ordered} operator form of $F(q,Q)$ (which contains $-2\hat q\hat Q$) does \textit{not} work, which is to say, it does not satisfy \Eqref{a5}. To see this in detail---and to see the cure---we assume the following small-$t$ limiting form for $\hat W$
\begin{equation}
\hat W= \frac m{2t}\left(\hat Q^2 -2\hat q\hat Q+\hat q^2\right)+g(t) \;.
\label{b1}
\end{equation}
Substituting into \Eqref{a5}, the squaring of $\hat W$ generates a term $-(\hat q\hat Q+\hat Q\hat q)$, rather than $-2\hat q\hat Q$, so that satisfying \Eqref{a5} requires
\begin{equation}
0= \frac m{2t^2}[\hat q, \hat Q]+\frac{\partial g(t)}{\partial t} \;.
\label{b2}
\end{equation}
For small $t$, one can again neglect the influence of the potential terms and the commutator can immediately be deduced from the relation $\hat q=\hat Q+\hat Pt/m$, the solution of the free particle Heisenberg equations of motion. \Eqref{b2} now becomes $\partial g/\partial t=i\hbar/2t$ and we obtain
\begin{eqnarray}
\hat W&=& \frac m{2t}\left(\hat Q^2 -2\hat q\hat Q+\hat q^2\right)+\frac {i\hbar}2 \, {\rm ln} \, t \;,
\qquad\hbox{for}\ t\to0 \;,
\\
\psi (q,Q,t)&=&\mathrm{const}\cdot\sqrt{\frac1t}\exp\left(\frac{i}\hbar \frac{m}{2t}\left(Q^2-2qQ-q^2\right)\right)
\;,\qquad\hbox{for}\ t\to0 \;,
\label{b3}
\end{eqnarray}
with the constant in \Eqref{b3} arising from integrating $\partial g/\partial t=i\hbar/2t$. It is remarkable that, aside from the constant (which is not fixed by $\hat W$), the ${\hat q}{\hat Q}$-commutation relation has given us precisely the correct time-dependence of the propagator. This completes our proof.

The c-number QHJE (\ref{a18}) has repeatedly attracted interest. For instance, \Eqref{a18} has been derived from a diffeomorphic covariance principle based partly on an $SL(2,C)$ algebraic symmetry of a Legendre transform~\cite{Faraggi}. Alternatively, \Eqref{a18} has been taken as the starting point of a classical-like strategy to define {\it c-number} quantum action-angle variables in quantum mechanics~\cite{Leacock}. We also remark that a variant of \Eqref{a18} has been derived within the phase-space path-integral approach to quantum mechanics~\cite{Periwal}.

\def\wo{\hbox{\tiny WO}}\def\barwo{|_{\wo}}

We next show the power of the relation we have just developed between $W(q, Q,t)$ and $K(q,Q,t)$, using, as suggested above, known information about the propagator. Consider a situation where the semiclassical approximation is valid and there is but one classical path between the initial and final points. Then in this approximation, as is well-known, 
$K(q,Q,t)=\mathrm{const}\cdot \sqrt{\det\partial^2S/\partial q\partial Q}\exp(iS(q,Q,t)/\hbar)$, with $S(q,Q,t)$ Hamilton's principal function (a solution of the \textit{classical} Hamilton-Jacobi equation). It then follows from our result that $W(\hat q,\hat Q,t)|_{\wo} = S(\hat q,\hat Q,t)\barwo-\frac12i\hbar \log\det\partial^2\hat S/\partial q\partial Q\barwo$, where ``WO'' stands for ``well-ordered.'' Now imagine that this expression is inserted in \Eqref{a5}. If not for the well ordering, $S$ alone would solve the equation. Therefore we conclude that the effect of the well-ordering is precisely to demand the presence of the additional term, $\frac12i\hbar \log\det\partial^2S/\partial q\partial Q$ (where ``WO'' has been dropped because there is already an $\hbar$ in the expression). But that additional term (famously) has a meaning of its own: it goes back to van Vleck and represents the density of paths along the classical path; it plays an essential role, for example, in the Gutzwiller trace formula. What our result says is that this density of paths can be thought of as arising from the commutation operations necessary to bring $S$ to well-ordered form. Thus the purely quantum issue of commuting operators produces a quantity that one would have thought is exclusively derivable from classical mechanics.

We remark that this relation took us completely by surprise. To check it, we worked the simplest non-trivial example we could (our proof above comparing the boundary conditions for $K$ and $W$ already showed it to be true for the free particle case). Let $H=p^2/2+V$ with $V=V_0\Theta(a/2-|x|)$ and $x$ in one dimension. To lowest order in $V$ the action is $S(x,y,t)=(x-y)^2/2t-V_0at/(x-y)$ for $y<-a/2$ and $x>a/2$. We checked our relation, with $x\to\hat q$ and $y\to\hat Q$ and with well-ordering implemented by $\left[1/(\hat q-\hat Q)\right]_{\wo}=\int_0^\infty du\,\exp(-u\hat q)\exp(u\hat Q)$. Using the Baker-Campbell-Hausdorff formula and other techniques and keeping only lowest order in $V$ and $\hbar$, indeed the relation checked out!

In conclusion, we have shown how to construct solutions to the operator quantum QHJE starting from the quantum propagator $K(q,Q,t)$ for the same Hamiltonian. Explicitly, once $K(q,Q,t)$ is known we get its ``complex phase'' $W(q, Q,t)$ via Eq. (\ref{a19}). Then, by demanding {\it 
well-ordering}, the replacement $q \to \hat q$, $Q \to \hat Q$ uniquely produces the operator $W(\hat q, \hat Q, t)$. Alternatively, by convolving $K(q,Q,t)$ with an arbitrary $\phi(Q)$ we produce any solution of the Schr\"odinger equation. Finally, by replacing $q \to \hat q$ in its ``complex phase'' we get a solution $W(\hat q,t)$ of the operator QHJE. While this is obviously true for exact propagators, it also enables one to find approximate solutions to the operator QHJE by exploiting approximate propagators. In particular we used the semiclassical approximation to the propagator to show that the commutation operations establishing well-ordering provide just what is needed to get the density of paths around the classical path. This density of paths satisfies a continuity equation which, as O'Raifeartaigh and Wipf \cite{wipf} emphasize, is in a sense of order $\hbar$ (even though it involves classical quantities only and has no $\hbar$ in it!). Although our proof establishes this surprising relation, there remains the provocative question of understanding its intuitive basis.

\bigskip\noindent\textbf{Acknowledgements.~} We thank the Max Planck Institute for the Physics of Complex Systems, Dresden, for kind hospitality. This work was supported by NSF \hbox{Grant PHY 0555313.}


\begin{thebibliography}{99}

\bibitem{Goldstein} H. Goldstein, {\it Classical Mechanics}, 2$^{\mathrm nd}$ ed.\ (Addison-Wesley, Reading, Mass.\ 1966).

\bibitem{Born} M. Born, W. Heisenberg and P. Jordan, Z. Phys.\ {\bf 35}, 557 (1926). P. A. M. Dirac, Proc.\ Roy.\ Soc.\ A{\bf 113}, 621 (1927).

\bibitem{Jordan} P. Jordan, Z. Phys.\ {\bf 37}, 383 (1926); {\bf 38}, 513 (1926).

\bibitem{London} F. London, Z. Phys. {\bf 37}, 915 (1926).

\bibitem{Dirac} P. A. M. Dirac, Physik.\ Zeits.\ Sowjetunion {\bf 3}, 64 (1933).

\bibitem{Schwinger1} J. Schwinger, Phys. Rev. {\bf 82}, 914 (1951); Phys. Rev. {\bf 91},
713 (1953).

\bibitem{Schwinger2} J. Schwinger, {\it Quantum Kinematics and Dynamics} (Benjamin, New York, 1970).

\bibitem{hat} As usual, quantum operators representing dynamical variables are denoted by a hat.

\bibitem{Sudarshan} T. F. Jordan and E. C. G. Sudarshan, Rev.\ Mod.\ Phys.\ {\bf 33}, 515 (1961).

\bibitem{cnumber} We follow Dirac's terminology, according to which ordinary, i.\ e., commuting, variables are called ``c-numbers.''

\bibitem{TDLee} See, e.\ g., T. D. Lee, {\it Particle Physics and Introduction to Field Theory} (Harwood Acad.\ Pub., New York, 1981).

\bibitem{Gottfried} Verifying this assertion is a textbook exercise, one that is done for the case $a=1/4m$, $b=0$, $c=V$ in Eq.\ (12), \S$\;$8.2, of K. Gottfried, {\it Quantum Mechanics},
1$^{\mathrm st}$ ed.\ (Benjamin, New York, 1966).

\bibitem{Schulman} See, e.\ g., L. S. Schulman, {\it Techniques and Applications of Path Integration} (Wiley, New York, 1981; Dover, New York, 2005).

\bibitem{Faraggi} A. Faraggi and M. Matone, Int.\ J. Mod.\ Phys.\ A{\bf 15}, 1869 (2000).

\bibitem{Leacock} R. A. Leacock and M. Padgett, Phys.\ Rev.\ Lett.\ {\bf 50}, 3 (1983); Phys. Rev. D{\bf 28}, 2491 (1983).

\bibitem{Periwal} V. Periwal, Phys.\ Rev.\ Lett.\ {\bf 80}, 4366 (1998).

\bibitem{wipf} L. O'Raifeartaigh and A. Wipf, Found.\ Phys.\ {\bf18}, 307 (1988).

\end{thebibliography}
\end{document}